# Leveraging Reactant Entanglement in the Coherent Control of Ultracold Bimolecular Chemical Reactions

Adrien Devolder[1], Timur V. Tscherbul[2], and Paul Brumer[1]

[1]*Chemical Physics Theory Group, Department of Chemistry,
and Center for Quantum Information and Quantum Control,
University of Toronto, Toronto, Ontario, M5S 3H6, Canada*

[2]*Department of Physics, University of Nevada, Reno, NV, 89557, USA*

(Dated: May 15, 2025)

Entanglement is a crucial resource for achieving quantum advantages in quantum computation, quantum sensing, and quantum communication. As shown in this Letter, entanglement is also a valuable resource for the coherent control of the large class of bimolecular chemical reactions. We introduce an entanglement-enhanced coherent control scheme, in which the initial preparation of the superposition state is divided into two steps: the first entangles the reactants, and the second is responsible for coherent control. This approach can overcome the limitations of traditional coherent control of scattering caused by non-interfering pathways, known as satellite terms. By tuning the amount of entanglement between reactants, the visibility of coherent control in chemical reactions can be modulated and optimized. Significantly, there exists an optimal amount of entanglement, which ensures complete indistinguishability of the reaction pathways, maximizing the extent of control. This entanglement-enhanced coherent control scheme is computationally illustrated using the ultracold KRb + KRb reaction, where a perfect control over the parity of the product rotational states is achieved.

Entanglement, arguably the most striking and counterintuitive feature of quantum mechanics [1], manifests itself as a non-local quantum correlation between particles as famously revealed by Bell's inequalities [2–4] and their experimental tests [3]. In quantum information science and technology, entanglement serves as a key resource, which can be harnessed to perform quantum computation [5], sensing [6], and communication [7]. Thus far, the vast majority of experimental studies of entangled states of matter have been carried out with ultracold atoms and ions [8–12]. Examples include squeezed-spin and Greenberger-Horne-Zeilinger (GHZ) states, which enable quantum-enhanced metrology beyond the standard quantum limit [13]. Additionally, studies of entanglement generation and propagation have provided invaluable insights into quantum information dynamics and thermalization in quantum many-body systems out of thermal equilibrium [14].

Ultracold molecules have recently emerged as another powerful platform for quantum science [15–17]. Their numerous vibrational and rotational degrees of freedom interacting via strong and tunable electric dipolar interactions offer unique opportunities for storing and processing quantum information in novel ways inaccessible to atoms. Recent experimental advances in high-fidelity entanglement generation between ultracold molecules trapped in optical tweezers [18–20] open up a host of tantalizing opportunities at the interface of chemical physics and quantum information science, including questions such as: What is the relationship between entanglement and chemical reactivity? Can one harness the entanglement to help control ultracold molecular collisions and chemical reactions? While addressing these questions is essential for exploring the emerging connections between ultracold molecular physics, chemical physics, and quantum information science, their theoretical exploration is still in its early infancy.

Entanglement has recently been considered at the level of potential energy landscapes and transition states [21], opening the door to using entanglement calculations for detecting these states. Nevertheless, these studies considered static properties and non-entangled reactants. To exploit entanglement as a resource for controlling chemical reactions, it is necessary to prepare the reactants in entangled states with a tunable amount of entanglement and analyze their quantum dynamics. Entangled states have been briefly considered in previous work on coherent control, a methodology that relies on preparing reactant molecules in superpositions of internal states such as vibrational, rotational, or spin states and tuning the relative phases of these superpositions to manipulate interference between different reaction pathways [22, 23]. Gong *et al.* [24] qualitatively showed an improved control with a Bell state of indistinguishable molecules. We later explored using entangled states to control ultracold spin-exchange collisions of $O_2$ molecules [25]. Other restricted cases were examined in [26–28]. Entangled states in chemical reactions can also be used to test Bell's inequality with continuous measurement outcomes as illustrated by Li *et al.* [29]. However, no previous studies have explored the relationship between entanglement and chemical reactivity. As such, the potential role of entanglement as a quantum resource to actively enhance quantum control of bimolecular chemical reactions has remained elusive.

In this Letter, we explore how entanglement between the reactant molecules affects chemical reactivity in molecular collisions. We show that the primary role of entanglement is to maximize the weight of interfering pathways leading from the reactants to products.

This mechanism can be used to harness entanglement to facilitate quantum interference-based coherent control of bimolecular chemical reaction by eliminating control-limiting satellite terms. Significantly, we find that entanglement does not always enhance coherent control, and we uncover a subtle relationship between the concepts of entanglement concurrence and visibility. We exemplify our results using computations on a realistic model of the ultracold chemical reaction KRb + KRb → $K_2$ + $Rb_2$. Our results can be experimentally verified with ultracold KRb molecules in optical tweezers [18, 19, 30–32]

Consider a bimolecular chemical reaction $A + B \to C + D$. For both reactant molecules, two internal states (e.g., spin, rotational, or vibrational) $\{|0\rangle_{A,B}, |1\rangle_{A,B}\}$ are used to form superpositions. Each reactant can be visualized as a molecular qubit. Therefore, we will use the notation and terminology of quantum information, for example by representing the operations on reactant molecules as quantum circuits [see Fig. 1]. Initially, both molecules occupy their $|0\rangle$ states. The preparation of the reactant superposition is divided into two steps (see FIG. 1 (a)). The first step is an entangling operation $U_{ent}(\boldsymbol{\theta})$, which entangles the two reactant molecules before the chemical reaction and depends on a set of parameters $\boldsymbol{\theta}$. Next, a rotation gate $R_z(\beta)$ is applied to the first molecule, which changes the relative phase between $|01\rangle$ and $|10\rangle$ and is responsible for coherent control of the reaction. After these two steps, the reactant molecules are in the general bipartite quantum state:

$$|\Psi_{ini}\rangle = c_{00}(\boldsymbol{\theta})|00\rangle + c_{01}(\boldsymbol{\theta})|01\rangle \\ + c_{10}(\boldsymbol{\theta})|10\rangle e^{i\beta} + c_{11}(\boldsymbol{\theta})e^{i\beta}|11\rangle. \quad (1)$$

The degree of entanglement in this state can be quantified using the concurrence:

$$C_{ent} = |c_{00}(\boldsymbol{\theta})c_{11}(\boldsymbol{\theta}) - c_{01}(\boldsymbol{\theta})c_{10}(\boldsymbol{\theta})|, \quad (2)$$

This initial superposition creates four different scattering paths, which can interfere with one another. However, the symmetries of the reactive collision limit the interference between these paths. Only paths created by states with the same energies and the same projection of the internal angular momentum can interfere [33]. Unfortunately, these two conditions – equal energy and identical internal angular momentum projection – make it impossible for all four reactive paths to interfere simultaneously. In the best-case scenario, interference occurs only between the reactive paths originating from $|01\rangle$ and $|10\rangle$. The paths associated with $|00\rangle$ and $|11\rangle$ do not contribute to interference terms and are referred to as satellite terms [22, 23].

After preparing the reactant in the initial state $|\Psi_{ini}\rangle$ (1), the reactants collide and reaction occurs. Different quantities can be measured, such as the total reaction cross section or product state populations, the latter

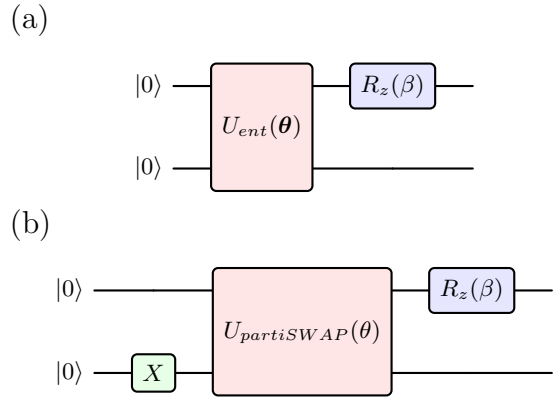

FIG. 1: (a) General quantum circuit for the entanglement-enhanced coherent control (b) Quantum circuit for the preparation of the initial state $|\Psi_{ini}\rangle$ (9)

being considered in the example of KRb-KRb chemical reaction later below. The reaction cross section is given by:

$$\sigma_{reac}(\boldsymbol{\theta}, \beta) = |c_{00}(\boldsymbol{\theta})|^2 \sigma_{00} + |c_{01}(\boldsymbol{\theta})|^2 \sigma_{01} \\ + |c_{10}(\boldsymbol{\theta})|^2 \sigma_{10} + |c_{11}(\boldsymbol{\theta})|^2 \sigma_{11} \\ + 2|c_{01}(\boldsymbol{\theta})||c_{10}(\boldsymbol{\theta})||\sigma_{int}|\cos(\Phi(\sigma_{int}) + \beta), \quad (3)$$

where $\sigma_{01}, \sigma_{01}, \sigma_{10}$ and $\sigma_{11}$ are the reaction cross sections out of the $|00\rangle, |01\rangle, |10\rangle$ and $|11\rangle$ initial states of the reactants, $\sigma_{int} = |\sigma_{int}|e^{i\Phi(\sigma_{int})}$ is the interference term (where $\Phi(\sigma_{int})$ is its argument) and is induced by the interference between the scattering paths created by the degenerate states $|01\rangle$ and $|10\rangle$. They are defined via the S-matrix elements:

$$\sigma_{ij} = \frac{\pi}{k^2} \sum_f \sum_{\ell m_\ell} \sum_{\ell' m'_\ell} \left| S_{ij,\ell,m_\ell \to f, \ell', m'_\ell} \right|^2, \quad (4)$$

$$\sigma_{int} = \frac{\pi}{k^2} \sum_f \sum_{\ell m_\ell} \sum_{\ell' m'_\ell} S_{01,\ell,m_\ell \to f, \ell', m'_\ell} S^*_{10,\ell,m_\ell \to f, \ell', m'_\ell}, \quad (5)$$

where $ij = \{00, 01, 10, 11\}$. The sum $\sum_f$ is over all the product states of interest. $\ell$ ($\ell'$) is the initial (final) partial wave, $m_\ell$ ($m'_\ell$) are the corresponding projection quantum numbers, and $S_{ij\ell m_\ell \to f\ell' m'_\ell}$ are the S-matrix elements from the initial two-molecule state $|ij\rangle$.

The difference, compared to the reaction cross section obtained from a statistical mixture $\rho = |c_{00}|^2 |00\rangle\langle 00| + |c_{01}|^2 |01\rangle\langle 01| + |c_{10}|^2 |10\rangle\langle 10| + |c_{11}|^2 |11\rangle\langle 11|$, is the term arising from interference between the paths created by the states $|01\rangle$ and $|10\rangle$. Quantum control of the reaction cross section is possible through this interference term. The purpose of the first entangling steps in the initial preparation of reactants is to enhance the weight of the interference terms relative to the others

(which could be produced by a statistical mixture) and hence to optimize the effectiveness of the coherent control.

The performance of coherent control is quantified by the visibility, defined as the ratio between the difference and the sum of the maximum and minimum values of the reaction cross section as the relative phase $\beta$ is varied:

$$V(\boldsymbol{\theta}) = \frac{\sigma_{reac}^{max}(\boldsymbol{\theta}) - \sigma_{reac}^{min}(\boldsymbol{\theta})}{\sigma_{reac}^{max}(\boldsymbol{\theta}) + \sigma_{reac}^{min}(\boldsymbol{\theta})} \\ = \frac{2R_c |c_{01}(\boldsymbol{\theta})||c_{10}(\boldsymbol{\theta})|\sqrt{\sigma_{01}\sigma_{10}}}{\sum_{i=0}^{1}\sum_{j=0}^{1} |c_{ij}(\boldsymbol{\theta})|^2 \sigma_{ij}}. \quad (6)$$

Here the control index $R_c$ quantifies the maximum achievable visibility and is defined via the Cauchy–Schwarz inequality between the interference term $\sigma_{int}$ and the geometric mean of the individual contributions $\sigma_{01}$ and $\sigma_{10}$: $R_c = |\sigma_{int}|/\sqrt{\sigma_{01}\sigma_{10}}$. This index reflects differences in the control landscapes associated with various product states, as well as differences due to various initial and final partial waves. When the control parameters influencing each product state and partial waves are well synchronized, $R_c$ approaches 1, enabling complete destructive interference. Conversely, when these controls are not well aligned, interference is reduced due to a competition among them, resulting in a lower value of $R_c$. In the case where only a single product state is measured, this reduction corresponds to the partial wave scrambling [27]. The value of the control index is determined by the S-matrix elements, and therefore depends on the properties of the collisional system, the chosen internal states of the reactants $\{|0\rangle_{A,B}, |1\rangle_{A,B}\}$, and the targeted product states. It can be enhanced, for instance, by using reactant internal states related through time-reversal symmetry [28], or by lowering the collisional energy [25].

The approach of entanglement-enhanced coherent control is to use an entangling operation to prepare an initial superposition that maximizes the visibility associated with the coherent control of the scattering event. *The entangling step enables coherent control to reach its full potential.* That is, a major limitation of the coherent control of scattering arises from the reactive paths created by $|00\rangle$ and $|11\rangle$ which neither interfere with one another nor with the paths originating from $|01\rangle$ and $|10\rangle$. This "satellite term" issue in coherent control can be resolved by preparing a superposition involving only the $|01\rangle$ and $|10\rangle$ states:

$$|\Psi_{ent}\rangle = \cos\theta |01\rangle - i\sin\theta |10\rangle. \quad (7)$$

Starting from both reactant molecules in their $|0\rangle$ states, the state in (7) is obtained [see Fig. 1 (b)] by first exciting the second molecule, $|00\rangle \to |01\rangle$, followed by the application of a partial iSWAP rotation gate, for which the implementation requires evolution under an interaction Hamiltonian of the exchange type: $H_{int} =$

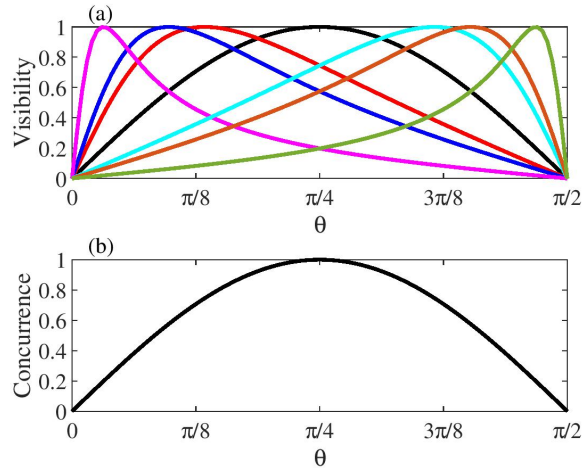

FIG. 2: (a) Visibility of coherent control as a function of the parameter $\theta$ of the initial state $|\Psi_{\text{ini}}\rangle$ (Eq. 9), shown for various values of the cross section ratio $\frac{\sigma_{01}}{\sigma_{10}}$: 0.01 (magenta), 0.1 (blue), 0.2 (red), 1 (black), 5 (cyan), 10 (orange), and 100 (green). The control index $R_c$ is assigned a value of 1. (b) Entanglement of the initial state $|\Psi_{\text{ini}}\rangle$ (Eq. 9), also as a function of $\theta$, quantified by the concurrence.

$J\left(\sigma_+^{(1)}\sigma_-^{(2)} + \sigma_-^{(1)}\sigma_+^{(2)}\right)$, where $\sigma_\pm^{(j)} = \frac{1}{2}\left(\sigma_x^{(j)} \mp \sigma_y^{(j)}\right)$ are the rising and lowering Pauli operators of the $j$-th molecular qubit, and $J$ is the interaction strength. The exchange interaction can be realized through the dipolar interaction between two polar molecules [19]. In this case, the interaction strength depends on the distance between the two molecules and their relative orientation. The evolution operator of this exchange interaction $e^{iH_{int}t}$ corresponds to the following unitary matrix:

$$U_{partiSWAP}(t) = e^{iH_{int}t} = \begin{pmatrix} 1 & 0 & 0 & 0 \\ 0 & \cos(Jt) & -i\sin(Jt) & 0 \\ 0 & -i\sin(Jt) & \cos(Jt) & 0 \\ 0 & 0 & 0 & 1 \end{pmatrix}. \quad (8)$$

By setting the interaction time $T = \frac{\theta}{J}$, this unitary operator $U_{partiSWAP}(T)$ prepares the state $|\Psi_{ent}\rangle$ in (7).

After the partial iSWAP rotation gate, a relative phase between $|01\rangle$ and $|10\rangle$ is introduced by applying a rotation gate $R_z(\beta)$ to the first molecule, enabling coherent control:

$$|\Psi_{ini}\rangle = \cos\theta |01\rangle - i\sin\theta e^{i\beta} |10\rangle. \quad (9)$$

The amount of entanglement in this state $|\Psi_{ini}\rangle$ (9) is determined by $\theta$:

$$C_{ent} = 2\sin\theta\cos\theta. \quad (10)$$

This dependence is illustrated in FIG. 2(b), which shows that maximal entanglement is achieved for $\theta = \pi/4$, corresponding to the preparation of the Bell state $\frac{1}{\sqrt{2}}\left(|01\rangle - ie^{i\beta}|10\rangle\right)$.

Using the prepared initial state $|\Psi_{ini}\rangle$ (9), coherent control of the scattering event yields the following visibility:

$$V(\theta) = R_c \frac{2\sin\theta\cos\theta\sqrt{\sigma_{01}/\sigma_{10}}}{\cos^2\theta(\sigma_{01}/\sigma_{10}) + \sin^2\theta}. \quad (11)$$

The $\theta$-dependence of the visibility $V$ is governed by the cross section ratio $\sigma_{01}/\sigma_{10}$ between the two reactive paths, as illustrated in Fig. 2(a). The maximum visibility ($V = R_c$) is achieved when $\theta$ satisfies:

$$\theta_{max} = \arctan\left(\sqrt{\sigma_{01}/\sigma_{10}}\right). \quad (12)$$

This behavior can be interpreted as the entanglement step trying to render the reactive pathways indistinguishable. Optimal coherent control is attained when the two reaction paths cannot be distinguished from one another. In cases where an asymmetry exists between $\sigma_{01}$ and $\sigma_{10}$, the entanglement control compensates by allocating more population to the state associated with the smaller cross section. This redistribution equalizes the contributions of the two paths, thereby enhancing their indistinguishability.

The entanglement corresponding to maximum coherent control is given by:

$$C_{max} = 2\frac{\sqrt{\sigma_{01}/\sigma_{10}}}{\sigma_{01}/\sigma_{10} + 1}, \quad (13)$$

and depends on the imbalance between $\sigma_{01}$ and $\sigma_{10}$. *When such an imbalance exists, the best coherent control is not achieved with a maximally entangled state. Instead, there is a particular amount of entanglement that ensures the indistinguishability of the reactive paths and enables the most effective coherent control. Increasing entanglement beyond this point would actually make the two paths more distinguishable, reducing control performance.*

To illustrate the entanglement-enhanced coherent control of ultracold bimolecular chemical reaction, consider the paradigmatic reaction KRb+KRb → K$_2$ + Rb$_2$ which has been broadly studied experimentally [30–32, 34–36]. The nuclear spin degrees of freedom have been shown to act as spectators during this chemical reaction [35], allowing them to avoid the chaotic dynamics often observed in collisions between ultracold molecules [37]. As a consequence of this spectator status, the nuclear spin state of the product molecules can be simply obtained using angular momentum algebra and a change of basis with Clebsh-Gordan coefficients [31, 32, 35].

Consider then two KRb reactant molecules prepared in the initial superposition $|\Psi_{ini}\rangle$ (9) using the two nuclear spin states $|I^K = 4, m^K = -4, I^{Rb} = 3/2, m^{Rb} = 1/2\rangle \equiv |0\rangle$ and $|4, -3, 3/2, -1/2\rangle \equiv |1\rangle$ described in an experimental coherent control proposal [32]. Importantly, here we consider the entanglement between the two KRb reactants. This is different from the entanglement between the nuclear spins of the atoms K and Rb inside of the KRb molecules used in Ref. [31].

The K$_2$ (Rb$_2$) reaction products are composed of identical bosonic (fermionic) atoms, so their wavefunctions must be symmetric (antisymmetric) under permutation. This symmetry imposes constraints on the possible combinations of nuclear spin and rotational states. For the K$_2$ molecule, a symmetric (S) nuclear spin state must be associated with an even rotational state, and an antisymmetric (A) nuclear spin state must be associated with an odd rotational state. Conversely, for the Rb$_2$ molecules: a symmetric (S) nuclear spin state must be associated with an odd rotational state, and an antisymmetric (A) nuclear spin state must be associated with an even rotational state. Therefore, the total wave function of the nuclei is:

$$|\Psi^{tot}_{K_2,Rb_2}\rangle = |SS\rangle|eo\rangle + |AA\rangle|oe\rangle + |SA\rangle|ee\rangle + |AS\rangle|oo\rangle. \quad (14)$$

The symmetry of nuclear spin states can be inferred from the parity of the rotational states, as has been experimentally demonstrated [31]. This allows for the extraction of the populations in the states $|SS\rangle|eo\rangle$, $|AA\rangle|oe\rangle$, $|SA\rangle|ee\rangle$ and $|AS\rangle|oo\rangle$.

Theoretically, these populations are obtained via the rules of angular momentum algebra (see Supplementary material):

$$P_{SS,eo}(\theta,\beta) = \frac{1}{4}\left|\cos\theta - i\sin\theta e^{i\beta}\right|^2 = P_{AA,oe}(\theta,\beta), \quad (15)$$

$$P_{SA,ee}(\theta,\beta) = \frac{1}{4}\left|\cos\theta + i\sin\theta e^{i\beta}\right|^2 = P_{AS,oo}(\theta,\beta). \quad (16)$$

These populations are modified by tuning the amount of entanglement between the reactants, and the relative phase between the reactive paths. The visibility for each population is given by:

$$V(\theta) = 2\sin\theta\cos\theta = C_{ent}. \quad (17)$$

In this case, the visibility equals the concurrence and reaches its maximum value of 1 at maximal entanglement [see Fig. 3(b)]. The final-state populations obtained with all the initial population in $|01\rangle$ ($\theta = 0$) are equal to the ones with all initial population in $|10\rangle$ ($\theta = \pi/2$). Therefore, maximum coherent control is achieved with an equal population of $|01\rangle$ and $|10\rangle$ states, and so with the Bell state $|\Psi_{ini}\rangle = \frac{1}{\sqrt{2}}\left(|01\rangle - ie^{i\beta}|10\rangle\right)$ ($\theta = \pi/4$). Coherent control using this maximally-entangled state is illustrated in FIG. 3 (a). By tuning the relative phase $\beta$, one can switch from a situation in which the product molecules have different parity (for $\beta = \frac{\pi}{2}$) to one in which they have the same parity (for $\beta = \frac{3\pi}{2}$). This demonstrates that, through entanglement-enhanced coherent control, perfect control ($V = R_c = 1$) over the parity of the product rotational states can be achieved.

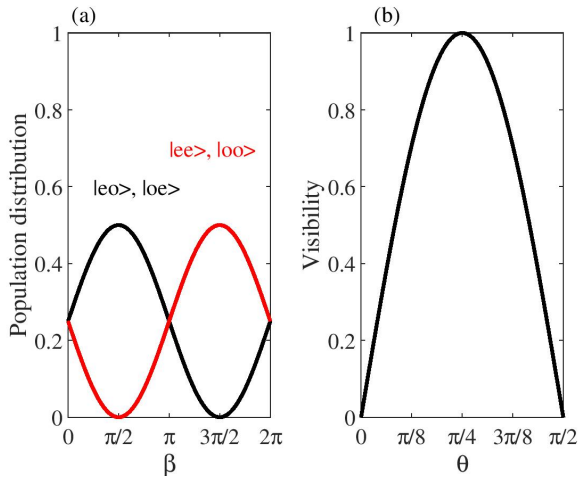

FIG. 3: (a) Coherent control of population distribution of products in the chemical reaction KRb+KRb $\to$ K$_2$ + Rb$_2$ using an initial maximally-entangled state $|\Psi\rangle = \frac{1}{\sqrt{2}} \left( e^{i\phi} |01\rangle - i |10\rangle \right)$. (b) Visibility for coherent control of population distribution as function of $\theta$ (related to the amount of entanglement)

In conclusion, we have elucidated the relationship between reactant entanglement and chemical reactivity by introducing a novel protocol of entanglement-assisted coherent control. As essential feature of this protocol is the initial entangling step, which allows coherent control to reach its full effectiveness. By preparing superposition states with population only in the $|01\rangle$ and $|10\rangle$ components, the entangling step resolves the issue of satellite terms that typically limit coherent control. We find a subtle relation between the visibility of coherent control and the amount of entanglement, parametrized by the imbalance between the two interfering reactive pathways. When this imbalance is present, optimal control is not obtained using a maximally entangled state, but a specific amount of entanglement is required to ensure the indistinguishability of the reactive pathways. We illustrated this entanglement–enhanced control approach using the ultracold KRb + KRb chemical reaction, demonstrating perfect control over the parity of the product rotational states, and paving the way toward efficient control of ultracold bimolecular chemical reactions through the combined use of entanglement and quantum interference. Note that entanglement-enhanced control is not limited to reactive scattering. Rather, it is applicable to a wide variety of processes involving interfering pathways, where two subsystems, defined by a partitioning of the initial Hilbert space, are used in the initial superposition. In such cases, the cross section noted in this letter is replaced by probabilities for the event of interest.

This work was supported by the U.S. Air Force Office of Scientific Research (AFOSR) under Contract No. FA9550-22-1-0361. SciNet computational facilities are gratefully acknowledged.

# Supplementary material: Harnessing Entanglement for Coherent Control of Ultracold Reactions

**PRODUCT STATE WAVEFUNCTION USING ANGULAR MOMENTUM ALGEBRA**

We define the reactant KRb nuclear spin state in the uncoupled basis of the K and Rb nuclear spins $|I^K, m^K, I^{Rb}, m^{Rb}\rangle$. Since nuclear spin states are conserved during the reaction, the nuclear rearrangement can be interpreted as a basis transformation from the uncoupled basis of both KRb molecules, $|I^K, m_1^K\rangle |I^K, m_2^K\rangle |I^{Rb}, m_1^{Rb}\rangle |I^{Rb}, m_2^{Rb}\rangle$, to the coupled basis of the $K_2$ and $Rb_2$ nuclear spins, $|I^{K_2}, m_I^{K_2}\rangle \otimes |I^{Rb_2}, m_I^{Rb_2}\rangle$. This transformation between the two bases is determined by the Clebsch–Gordan coefficients:

$$|I^K, m_1^K\rangle |I^K, m_2^K\rangle = \sum_{I^{K_2}} C_{I^K m_1^K I^K m_2^K}^{I^{K_2} m_1^K + m_2^K} |I^{K_2}, m_1^K + m_2^K\rangle \tag{1}$$

$$|I^{Rb}, m_1^{Rb}\rangle |I^{Rb}, m_2^{Rb}\rangle = \sum_{I^{Rb_2}} C_{I^{Rb} m_1^{Rb} I^{Rb} m_2^{Rb}}^{I^{Rb_2} m_1^{Rb} + m_2^{Rb}} |I^{Rb_2}, m_1^{Rb} + m_2^{Rb}\rangle \tag{2}$$

where $m_1^{K(Rb)}$ and $m_2^{K(Rb)}$ are the projection of the nuclear spin of the K (Rb) atoms in the first and the second KRb molecules, respectively. In K.K Ni experiment, the isotope 40 is used for the potassium atoms, while the isotope 87 is used for the rubidium atoms. Therefore, $I^K = 4$ and $I^{Rb} = 3/2$.

Two hyperfine states $|4, -4, 3/2, 1/2\rangle \equiv |0\rangle$ and $|4, -3, 3/2, -1/2\rangle \equiv |1\rangle$ are used for the superposition of the reactant KRb molecules. Note that only the states $|01\rangle$ and $|10\rangle$ interfere with each other. This can be explained by the sum of the projection for the potassium atoms, $m^{K_2} = m_1^K + m_2^K$ and by the sum of the projections for the rubidium atoms, $m^{Rb_2} = m_1^{Rb} + m_2^{Rb}$. To get the same coupled spin states $|I^{K_2}, m^{K_2}\rangle \otimes |I^{Rb_2}, m^{Rb_2}\rangle$ from the transformation (1) and (2), $m^{K_2}$ and $m^{Rb_2}$ must be the same. Then, in this case, we have:

$|00\rangle$: $m^{K_2} = -4 - 4 = -8$ and $m^{Rb_2} = 1/2 + 1/2 = 1$.
$|01\rangle$: $m^{K_2} = -4 - 3 = -7$ and $m^{Rb_2} = 1/2 - 1/2 = 0$.
$|10\rangle$: $m^{K_2} = -3 - 4 = -7$ and $m^{Rb_2} = -1/2 + 1/2 = 0$.
$|11\rangle$: $m^{K_2} = -3 - 3 = -6$ and $m^{Rb_2} = -1/2 - 1/2 = -1$.

This explain why only the states $|10\rangle$ and $|01\rangle$ interfere with each other. Therefore, the targeted initial state of the two KRb reactant molecules is a superposition of only these two states:

$$|\Psi_{ini}\rangle = \cos\theta |01\rangle - i\sin\theta e^{i\beta} |10\rangle \tag{3}$$

Using the transformation given by eqs. (1) and (2), we obtain the following final states for the product $K_2$ and $Rb_2$:

$$
\begin{aligned}
|\Psi^{nuc}_{K_2,Rb_2}\rangle = &-\frac{1}{2\sqrt{20}}(\cos\theta - i\sin\theta e^{i\beta})\,|8,-7\rangle\,|1,0\rangle + \frac{3}{2\sqrt{20}}(\cos\theta - i\sin\theta e^{i\beta})\,|8,-7\rangle\,|3,0\rangle \\
&+ \frac{1}{2\sqrt{4}}(\cos\theta - i\sin\theta e^{i\beta})\,|7,-7\rangle\,|0,0\rangle - \frac{1}{2\sqrt{4}}(\cos\theta - i\sin\theta e^{i\beta})\,|7,-7\rangle\,|2,0\rangle \\
&- \frac{1}{2\sqrt{4}}(\cos\theta + i\sin\theta e^{i\beta})\,|8,-7\rangle\,|0,0\rangle + \frac{1}{2\sqrt{4}}(\cos\theta + i\sin\theta e^{i\beta})\,|8,-7\rangle\,|2,0\rangle \\
&+ \frac{1}{2\sqrt{20}}(\cos\theta + i\sin\theta e^{i\beta})\,|7,-7\rangle\,|1,0\rangle - \frac{3}{2\sqrt{20}}(\cos\theta + i\sin\theta e^{i\beta})\,|7,-7\rangle\,|3,0\rangle
\end{aligned}
\quad (4)
$$

As explained in the main text, the permutation symmetry of the two K atoms and the two Rb atoms must be respected. This gives restriction on the possible states for the nuclear spin state and rotational states. For the $K_2$ molecule, a symmetric (S) nuclear spin state must be associated with an even rotational state, and an antisymmetric (A) nuclear spin state must be associated with an odd rotational state. It is the inverse for the $Rb_2$ molecules: a symmetric (S) nuclear spin state must be associated with an odd rotational state, and an antisymmetric (A) nuclear spin state must be associated with an even rotational state. Therefore, the total wave function of the nuclei are:

$$|\Psi^{tot}_{K_2,Rb_2}\rangle = |SS\rangle\,|eo\rangle + |AA\rangle\,|oe\rangle + |SA\rangle\,|ee\rangle + |AS\rangle\,|oo\rangle \quad (5)$$

When combining two spins $I_1$ and $I_2$ to form $I$, the resulting spin state will be symmetric (S) if $I - I_1 - I_2$ is even and antisymmetric (A) if $I - I_1 - I_2$ is odd. For $K_2$, $I_1 = I_2 = 4$, the coupled spin state is symmetric when $I^{K_2}$ is even, and is antisymmetric when $I^{K_2}$ is odd. For $Rb_2$, $I_1 = I_2 = 3/2$, the situation is opposite.

Therefore, using eq. (4), we obtain the $|SS\rangle$, $|AA\rangle$, $|SA\rangle$ and $|AS\rangle$ states:

$$|SS\rangle = -\frac{1}{2\sqrt{20}}(\cos\theta - i\sin\theta e^{i\beta})\,|8,-7\rangle\,|1,0\rangle + \frac{3}{2\sqrt{20}}(\cos\theta - i\sin\theta e^{i\beta})\,|8,-7\rangle\,|3,0\rangle \quad (6)$$

$$|AA\rangle = \frac{1}{2\sqrt{4}}(\cos\theta - i\sin\theta e^{i\beta})\,|7,-7\rangle\,|0,0\rangle - \frac{1}{2\sqrt{4}}(\cos\theta - i\sin\theta e^{i\beta})\,|7,-7\rangle\,|2,0\rangle \quad (7)$$

$$|SA\rangle = -\frac{1}{2\sqrt{4}}(\cos\theta + i\sin\theta e^{i\beta})\,|8,-7\rangle\,|0,0\rangle + \frac{1}{2\sqrt{4}}(\cos\theta + i\sin\theta e^{i\beta})\,|8,-7\rangle\,|2,0\rangle \quad (8)$$

$$|AS\rangle = \frac{1}{2\sqrt{20}}(\cos\theta + i\sin\theta e^{i\beta})\,|7,-7\rangle\,|1,0\rangle - \frac{3}{2\sqrt{20}}(\cos\theta + i\sin\theta e^{i\beta})\,|7,-7\rangle\,|3,0\rangle \quad (9)$$

Finally, we obtain the populations in $|SS\rangle\,|eo\rangle$, $|AA\rangle\,|oe\rangle$, $|SA\rangle\,|ee\rangle$ and $|AS\rangle\,|oo\rangle$ states given in the main text:

$$P_{SS,eo}(\theta,\beta) = \frac{1}{4}\left|\cos\theta - i\sin\theta e^{i\beta}\right|^2 = P_{AA,oe}(\theta,\beta), \quad (10)$$

$$P_{SA,ee}(\theta,\beta) = \frac{1}{4}\left|\cos\theta + i\sin\theta e^{i\beta}\right|^2 = P_{AS,oo}(\theta,\beta). \quad (11)$$